\documentclass{emulateapj}

\shorttitle{A Search for Companions in Taurus and Chamaeleon}
\shortauthors{Todorov et al.}
\begin{document}
\title{A Search for Companions to Brown Dwarfs in the Taurus and 
Chamaeleon Star Forming Regions\altaffilmark{1}}

\author{
K. O. Todorov\altaffilmark{2,3},
K. L. Luhman\altaffilmark{2,4},
Q. M. Konopacky\altaffilmark{5},
K. K. McLeod\altaffilmark{6},
D. Apai\altaffilmark{7,8},
A. M. Ghez\altaffilmark{9,10},
I. Pascucci\altaffilmark{7,8},
\& M. Robberto\altaffilmark{11,12}}

\altaffiltext{1}{Based on observations performed with the NASA/ESA
{\it Hubble Space Telescope}, Gemini Observatory, and the W. M. Keck Observatory.
The {\it Hubble} observations are associated with proposal IDs 11203, 11204,
and 11983 and were obtained at the Space Telescope Science Institute,
which is operated by the Association of Universities for Research in 
Astronomy, Inc., under NASA contract NAS 5-26555.}
\altaffiltext{2}{Department of Astronomy and Astrophysics, The Pennsylvania
State University, University Park, PA 16802}
\altaffiltext{3}{Current address: Institute for Astronomy, ETH Zurich,
Wolfgang-Pauli-Strasse 27, CH-8093 Zurich, Switzerland; todorovk@phys.ethz.ch.}
\altaffiltext{4}{Center for Exoplanets and Habitable Worlds, 
The Pennsylvania State University, University Park, PA 16802.}
\altaffiltext{5}{Lawrence Livermore National Laboratory, 
7000 East Avenue, Livermore, CA 94550.}
\altaffiltext{6}{Whitin Observatory, Wellesley College, Wellesley, MA 02481.}
\altaffiltext{7}{Department of Astronomy, University of Arizona,
933 N. Cherry Avenue, Tucson, AZ 85721, USA}
\altaffiltext{8}{Lunar and Planetary Laboratory, 1629 E. University Blvd.,
Tucson, AZ 85721, USA}
\altaffiltext{9}{University of California, Los Angeles, Division of Astronomy
and Astrophysics, Los Angeles, CA 90095.}
\altaffiltext{10}{Institute of Geophysics and Planetary Physics, University of
California, Los Angeles, CA 90095.}
\altaffiltext{11}{Space Telescope Science Institute, 3700 San Martin Drive
Baltimore, MD 21218.}
\altaffiltext{12}{Johns Hopkins University, Center for Astrophysical Sciences
3400 North Charles Street, Baltimore, MD 21218.}

\begin{abstract}

We present the results of a search for companions to young brown dwarfs in the
Taurus and Chamaeleon~I star forming regions ($\tau\sim1$ and 2--3~Myr).
We have used
the Wide Field Planetary Camera 2 on board the {\it Hubble Space Telescope}
to obtain F791W and F850LP images of 47 members of these regions that have
spectral types of M6--L0 ($M\sim0.01$--0.1~$M_\odot$).
An additional late-type member of Taurus, FU~Tau (M7.25+M9.25), was also
observed with adaptive optics at Keck Observatory.
We have applied PSF subtraction to the primaries
and have searched the resulting images for objects that have colors
and magnitudes that are indicative of young low-mass objects.
Through this process, we have identified promising candidate companions
to 2MASS J04414489+2301513 ($\rho=0\farcs105$/15~AU), 2MASS J04221332+1934392
($\rho=0\farcs05$/7~AU), and ISO~217 ($\rho=0\farcs03$/5~AU).
We reported the discovery of the first candidate in a previous study,
showing that it has a similar proper motion as the primary through
a comparison of astrometry measured with WFPC2 and Gemini adaptive optics.
We have collected an additional epoch of data with Gemini that further
supports that result.
By combining our survey with previous high-resolution imaging in Taurus,
Chamaeleon~I, and Upper Sco ($\tau\sim10$~Myr), we measure binary fractions
of 14/93 = $0.15^{+0.05}_{-0.03}$ for M4--M6 ($M\sim0.1$--0.3~$M_\odot$) 
and 4/108 = $0.04^{+0.03}_{-0.01}$ for $>$M6 ($M\lesssim0.1$~$M_\odot$)
at separations of $>10$~AU.
Given the youth and low density of these three regions, the lower
binary fraction at later types is probably primordial rather than
due to dynamical interactions among association members.
The widest low-mass binaries ($>100$~AU) also appear to be more common in
Taurus and Chamaeleon~I than in the field, 
which suggests that the widest low-mass binaries are disrupted by dynamical
interactions at $>$10~Myr, or that field brown dwarfs have been born
predominantly in denser clusters where wide systems are disrupted or
inhibited from forming.

\end{abstract}


\section{Introduction}

As with stars at higher masses, the binary properties of low-mass stars
and brown dwarfs may provide insight into their formation and dynamical
evolution \citep{duc13}. Multiplicity at low masses has been characterized
primarily through high-resolution imaging in the solar neighborhood
\citep{koe99,mar99,rei01,bou03,bur03,clo03} and the nearby young clusters
and associations \citep{neu02,mar03,luh05ic,kra05,kra06,kon07,bil11,kra12}.
These surveys have found that the binary fractions and the
separations of binaries decrease and the mass ratios increase
from stars to brown dwarfs \citep{bur07ppv,kra12}.
Although most binary brown dwarfs have small separations ($a<20$~AU),
a few wide systems have been uncovered \citep{luh04bin,cha04}.
The low binding energies of these wide binaries would tend to suggest that
dynamical interactions did not play a role in their formation \citep{rc01},
although a few ejected brown dwarfs may be captured into wide systems
in denser clusters \citep{bat05}.

The dependence of the binary properties of low-mass stars and brown dwarfs on
age and star-forming environment is not well-constrained by existing data.
As a result, it is unclear whether wide low-mass binaries are frequently
disrupted by interactions with stars, either in their natal clusters
or in the Galactic field, and how the initial conditions of star formation
influence binarity at low masses. The Taurus and Chamaeleon~I star-forming
regions are promising sites for providing some of the data needed to
investigate these issues. They are among the nearest star-forming regions
(140 and 165~pc), young enough that dynamical interactions are minimized
(1 and 2--3~Myr), most of their members have relatively low extinction
($A_V\lesssim4$), and they have been searched thoroughly for substellar
members \citep{ken08,luh08cha}.
These two regions also offer the opportunity for characterizing the multiplicity
produced by low-density star-forming conditions, which can be compared
to measurements in richer and denser clusters at the same age
as well as older populations in open clusters and the solar neighborhood.

High-resolution imaging has been previously applied to low-mass members
of Taurus and Chamaeleon~I \citep{neu02,kra06,ahm07,kon07,luh07cha,laf08,kra12},
which has included 45 primaries with spectral types later than M6
($M\lesssim0.1$~$M_\odot$).
To improve upon the statistical accuracy of those multiplicity measurements,
we have performed an imaging survey that includes most of the remaining
known late-type members of Taurus and Chamaeleon~I using the
{\it Hubble Space Telescope} ({\it HST}) and Keck Observatory,
resulting in a sample of 73 primaries in these regions later than M6 for which
high-resolution data are now available. In this paper, we describe the sample
selection and observing strategy in our survey (Section \ref{sec:obs}) and our
analysis of the resulting images (Section \ref{sec:images}).
We then identify the most promising candidate companions in our data
and combine our sample with those of previous surveys to 
measure the binary fraction as a function of spectral type
(Section~\ref{sec:cand}). We conclude by discussing the implications of
our survey for measurements of multiplicity at low masses and for
the formation of brown dwarfs (Section~\ref{sec:disc}).

\section{Observations}
\label{sec:obs}

\subsection{Hubble Images}

\subsubsection{Sample Selection}
\label{sec:sample}

We have obtained most of the data in our survey with the Wide Field
Planetary Camera 2 (WFPC2) on board {\it HST}.
In our proposal for WFPC2 imaging in Taurus,
we selected all members of this region that had spectral types later than M6,
that had not been previously imaged with {\it HST} \citep{kra06}, and that
were known as of early 2007, which corresponded to 32 objects. 
However, observations could not be scheduled for three of the targets,
2MASS J04325026+2422115, 2MASS J04335245+2612548\footnote{This object has been
observed with adaptive optics (AO) imaging \citep{kra12}.}, and
2MASS J04380083+2558572\footnote{This object has been observed with
speckle imaging \citep{kon07}.}.
We replaced these sources with two Taurus members that have spectral types
of M6 (2MASS J04350850+2311398, 2MASS J04400067+2358211) and a new
member later than M6 that we uncovered after submission of the original
proposal \citep[2MASS J04373705+2331080,][]{luh09tau}.
Our WFPC2 images of these 32 targets
encompassed four additional members of Taurus, FM~Tau, V773~Tau, CW~Tau,
and 2MASS J04414565+2301580, all of which were saturated.
One of our targets has been previously observed with speckle imaging
\citep[2MAS J04442713+2512164,][]{kon07}. That object and 18 other targets have
been observed with AO imaging \citep{kra12}.

For the WFPC2 sample in Chamaeleon~I, we selected all known members
that are later than M6 and that have not been previously observed with
{\it HST} \citep{neu02,luh07cha} with the exception of
2MASS J11195652-7504529 and 2MASS J11070369-7724307, which were omitted
because the former is in the outskirts of the cluster and the latter is
highly reddened. The resulting sample contained 19 objects.
The observations of one target failed (ESO H$\alpha$ 554) and
three targets were not observed before the decommissioning of WFPC2
(2MASS J11085176-7632502, 2MASS J11104006-7630547,
ISO~138\footnote{This object has been observed with AO imaging
\citep{laf08}.}). Thus, we were able to obtain WFPC2 images of 15 late-type
primaries in Chamaeleon~I.
OTS~32, C1-2, 2MASS J11011926-7732383~B, CHXR~84, and Hn11 also fell within
the field of view of these data. The latter two stars were saturated.
One of our 15 targets, CHSM~17173, has been observed with AO imaging \citep{laf08}.

We present in Table~\ref{tab:st} the sample of 47 late-type primaries in
Taurus and Chamaeleon~I that we have observed with WFPC2. We also include
the extra members that appeared within the images and that
were not saturated (OTS~32, C1-2, 2MASS J11011926-7732383~B) as well
as the new companion to 2MASS J04414489+2301513 (hereafter 2M~J044144) that we
found in this survey and reported in an earlier study \citep{tod10}.

To obtain high-resolution images for two late-type members of Chamaeleon~I
that have high extinctions and thus were inaccessible with WFPC2, we
used the Near-Infrared Camera and Multi-Object Spectrometer (NICMOS)
on board {\it HST}. These objects consisted of 2MASS J11070369-7724307 (M7.5)
and 2MASS J11062942-7724586 (M6).

\subsubsection{Observing Strategy}

We obtained the WFPC2 images through the F791W and F850LP
filters, which are similar to the Cousins $I$ and SDSS $z\arcmin$ filters,
respectively. We selected these filters because they provide the optimum
combination of sensitivity to cool companions and spatial resolution 
and they produce a color-magnitude diagram that is effective in
distinguishing young low-mass objects from field stars.
WFPC2 contained four $800\times800$ CCDs. The plate scales of the
PC and the three WFC arrays were $0\farcs046$~pixel$^{-1}$ and
$0\farcs1$~pixel$^{-1}$, respectively.
To mitigate the effects of degraded charge-transfer efficiency (CTE),
we placed each target near the readout amplifier in the PC array.
Each target was observed with a two-point dither pattern.
At a given dither position, we obtained two images in each of the two filters.
We selected three combinations of exposure times and gains for three ranges
of optical magnitudes in order to avoid saturation of the targets.
From the faintest to brightest
targets, we used 1) $\tau_{791}=260$~s, $\tau_{850}=160$~s, gain=7,
2) $\tau_{791}=260$~s, $\tau_{850}=160$~s, gain=15, and
3) $\tau_{791}=200$~s, $\tau_{850}=160$~s, gain=15.
These exposure times apply to each of the four individual images for a given
filter. For the third set of exposure times, we also included a pair
of dithered images with exposure times of 40~s in F791W to provide unsaturated
data for the center of the point spread function (PSF). We did not specify the
position angle of the camera on the sky for these observations. Each object was
observed during one orbit, corresponding to a total of 47 orbits.

The NICMOS observations were performed with the NIC2 camera and the
F110W, F160W, and F205W filters. The camera contained a $256\times256$ array 
that had a plate scale of $0\farcs075$~pixel$^{-1}$.
We obtained one image in each filter at each position
in a 6-point dither pattern that was centered in the NIC2 array.
The exposure times of the individual images were 128, 128, and 96 sec
in F110W, F160W, and F205W, respectively.
Immediately before and after the dither sequence on each target, we 
collected one image in F205W at a position 30$\arcsec$ from the target
to measure the background. Each of the two NICMOS targets was observed
during one orbit.

\subsection{Keck Images}

Additional late-type members of Taurus have been uncovered since the
planning of our {\it HST} observations. We obtained high-resolution
images of two of these objects, FU~Tau~A and B \citep{luh09fu}, with the
near-infrared camera NIRC2 (PI: K. Matthews) in conjunction with the
laser guide star AO system at the Keck II 10~m telescope
\citep{wiz06,van06}. These observations were performed on 2008 December 18 and
2010 December 9. The target was bright enough to provide the tip/tilt
correction. The plate scale of NIRC2 is 9.952$\pm$0.002~mas~pixel$^{-1}$ and
its columns are rotated by $0.252\pm0.009\arcdeg$ relative to the nominal
position angle computed from the image headers \citep{yel10}.
We collected three $H$-band images at each position in a 3-point dither pattern.
The individual images consisted of 10 and 50 coadditions of 0.5~sec exposures
during the observations in 2008 and 2010, respectively.

\subsection{Gemini Images}

Our analysis of the WFPC2 data in Section~\ref{sec:psffit} reveals a
candidate companion to 2M~J044144.  In \citet{tod10}, we obtained
AO images of this pair to better constrain the nature of the candidate.
The images were taken in $H$ and $K\arcmin$ filters with the Gemini
Near-Infrared Imager \citep[NIRI,][]{hod03} and the ALTAIR AO system
at the Gemini North telescope.  The plate scale was $0\farcs0214$~pixel$^{-1}$
and the field of view was $22\arcsec\times22\arcsec$.
The relative positions of 2M~J044144 and its
candidate companion remained unchanged between the WFPC2 and AO observations,
indicating that the candidate shares a similar proper motion as the
primary and thus is a Taurus member rather than a field star.
The tip-tilt star for the AO observations was 2MASS J04414565+2301580
(hereafter 2M~J044145), which
is a member of Taurus that may be a wide companion to 2M~J044144 ($12\farcs3$).
The AO images from \citet{tod10} resolved a faint candidate companion at
$0\farcs23$ from that star, which could make 2M~J044144 and 2M~J044145 a
quadruple system.
To assess this possibility, we sought constraints on the relative
proper motions of the four objects by observing them again with NIRI+ALTAIR.
The data were collected on the night of 2011 February 13. The observing
strategy was the same as for the first epoch from \citet{tod10} except that one
dither sequence was performed in each filter instead of two.

\section{Image Analysis}
\label{sec:images}

\subsection{Reduction of Hubble Images}

We employed the MultiDrizzle software package \citep{2002mdrz} for
performing cosmic ray rejection on the WFPC2 images and for combining
the dithered frames for a given filter and exposure time.
We adopted a drop size of 0.85 native pixels
and resampled plate scales of $0\farcs01$~pixel$^{-1}$
and $0\farcs05$~pixel$^{-1}$ for the PC and WFC images, respectively.
Using the IRAF routine {\it starfind}, we identified all point sources
in each of the reduced PC and WFC images. Spurious detections
were manually removed through visual inspection of the images.
To search for sources that are blended with the primaries targeted by our
survey, we applied PSF subtraction, as described in the next section.

We measured aperture photometry for all unsaturated point sources
using the IRAF task {\it phot} with an aperture radius of 2 pixels on
the native scale, corresponding to $0\farcs091$
and $0\farcs2$ for the PC and WFC images, respectively.
We estimated aperture corrections between these radii and an aperture of
$0\farcs5$ for each array and filter using stars that were
isolated, bright, and unsaturated. The average values were
0.22 (F791W/WFC), 0.24 (F850LP/WFC), 0.54 (F791W/PC), and 0.61~mag (F850LP/PC).
The aperture correction from $0\farcs5$ to an infinite aperture is 0.1~mag. 
We arrived at the final photometric magnitudes by combining the 
measurements from {\it phot} with the aperture corrections, CTE 
corrections\footnote{http://purcell.as.arizona.edu/wfpc2\_calib},
and the zero point fluxes from the image headers.
The CTE corrections were larger for fainter objects. Since all of the
primaries were well-detected, their CTE corrections were fairly small
($\lesssim0.1$~mag). However, because of the advanced age of WFPC2
at the time of our observations, the CTE corrections were quite large for
the faintest objects (0.5--1~mag).
The magnitude at which saturation occurs is brighter in Taurus than in
Chamaeleon I because short exposures were included for some of the former
targets.  The photometry for the known members of Taurus and Chamaeleon~I
within the WFPC2 images is provided in Table~\ref{tab:st}. The uncertainties
in these measurements are $\sim0.05$~mag, which are dominated by the
errors in the aperture corrections.

As with the WFPC2 data, we reduced the NICMOS images with MultiDrizzle.
The resampled plate scale was $0\farcs025$~pixel$^{-1}$.
The field of view of each reduced image was small enough
($22\arcsec\times22\arcsec$) that visual inspection was adequate for
identifying sources in these images.
In the F110W images, the only detected objects consisted of
the two members of Chamaeleon~I that were targeted.
For both F160W and F205W, two and four additional sources were detected in
the images of 2MASS~J11070369-7724307 and 2MASS J11062942-7724586, respectively.
One of these objects is Cha~J11062788$-$7724543, which is $7\arcsec$
from 2MASS J11062942-7724586 and has been identified as a candidate low-mass
protostar based on its red mid-infrared (IR) colors \citep{luh08cha1}. The other
sources have separations of $>3\arcsec$ from the primaries. They are
unlikely to be cluster members based on photometry from other telescopes
\citep{luh07cha,luh08cha1}. Our analysis
to check for marginally resolved companions is described in the next section.

\subsection{PSF Analysis of Hubble Images}
\label{sec:psffit}

To detect sources at small angular separations from the survey primaries
in the WPFC2 and NICMOS images, 
we performed PSF subtractions using the IMFITFITS software, written by
Brian McLeod and described in \citet{leh00}.
For each primary, we fit its PSF with the PSF of every other primary that
was observed in that star-forming region. We then visually inspected each
of the subtracted images and identified the ones with the smallest residuals.
The PSFs of other primaries were better choices for PSF subtraction
than other stars in the image of a given primary because all of the primaries
have similar colors and were observed at the same location on the PC array.
Synthetic PSF fitting (e.g., TinyTim) was not attempted since it provides 
poorer fits than observed stellar PSFs \citep{luh05ic}. 

\begin{figure}
\epsscale{1}
\plotone{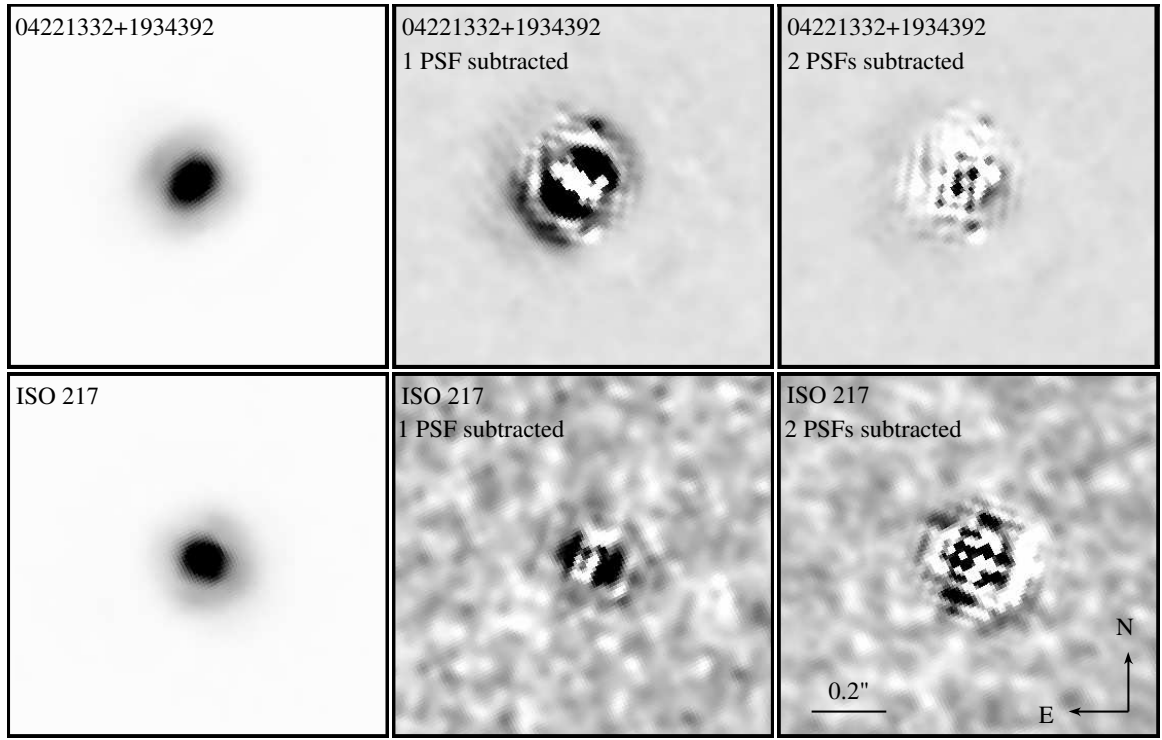}
\caption{
WFPC2 F791W images of the young late-type objects 2MASS J04221332+1934392 and
ISO 217 before and after PSF subtraction. Large symmetric residuals remain
after subtraction of single PSFs for both objects, indicating that they
are marginally resolved binaries. The residuals are much smaller when
each image is fit with a pair of PSFs. For each image prior to PSF subtraction,
the maximum of the intensity scale is 50\% of the peak of the PSF.
The upper limits for the scales in the PSF-subtracted images are
2\% and 1\% of the PSF peaks for 2MASS J04221332+1934392 and ISO 217,
respectively.
}
\label{fig:close}
\end{figure}

We found significant residuals after subtraction of a single PSF to
2M~J044144, 2MASS~J04221332+1934392, and ISO~217.
PSF subtraction for 2M~J044144 revealed a faint companion
at a separation of $0\farcs105$, as described by \citet{tod10}.
The other two objects appear to be marginally resolved binaries.
To measure the relative positions and fluxes of the components of each system,
we created a grid of subtracted images that used pairs of PSFs 
with a range of separations, position angles, and flux ratios.
Through visual inspection of the residuals in this grid of subtracted images,
we estimated the binary parameters and their uncertainties. These measurements
are presented in Table~\ref{tab:bin}.
Figure~\ref{fig:close} shows the images of 2MASS~J04221332+1934392
and ISO~217 produced by subtraction of the best-fit single and double PSFs.

\subsection{Reduction and PSF Analysis of Keck Images}

The Keck NIRC2 data for FU~Tau~A and B were processed using standard reduction
techniques for near-IR images. For each image, we subtracted a frame from
another dither position to remove the sky background, applied a mask for bad
pixels, divided by a flat field image, and corrected for optical distortion
with a model provided in the pre-ship review
document\footnote{http://www2.keck.hawaii.edu/inst/nirc2/preship$\_$testing.pdf}
using IRAF and IDL routines. The dithered frames were then registered and
combined.

Given their separation of $5\farcs7$, FU~Tau~A and B were well-resolved
from each other in the NIRC2 images.  Each component appears unresolved
in these data without an obvious additional companion.
To check for marginally resolved companions,
we applied the IDL package StarFinder to FU~Tau~A and B \citep{dio00},
which has a deblend function designed to detect
close pairs given a good empirical PSF. We used each component as the
PSF for the other object. In addition, separate
single sources were observed close enough in time to FU Tau A
that we could use them as secondary checks of the results of StarFinder.  
Our StarFinder analysis of the 2008 data showed a secondary source at a
separation of $\sim$2.5 pixels (or $\sim$0$\farcs$025) and a position angle
of $\sim9\arcdeg$ from FU~Tau~A with a flux ratio of $\sim$2.
In the images from 2010, no source was detected
at a comparable separation, which could be due to the difference in image
quality on the two nights.  The observing conditions during the
night in 2008 were very good, with a Strehl ratio estimated from the PSF of
$\sim20$\% in the $H$ band.  The conditions in 2010 were worse,
yielding a Strehl ratio of only 5$\%$.
Additional observations are necessary to determine definitively whether
FU~Tau~A is a binary as implied by the data from 2008.
We treat it as unresolved for the purposes of this study.

Our NIRC2 data have provided relative positions for the
components of FU~Tau that are more accurate than previous measurements from
seeing-limited ground-based images.
We measured a separation of $5.69\pm0.02\arcsec$ and
$5.69\pm0.01\arcsec$, a position angle of $122.75\pm0.24\arcdeg$ and
$122.77\pm0.02\arcdeg$, and an $H$-band flux ratio of $21\pm16$ and $35\pm3$
in the 2008 and 2010 images, respectively. These astrometric measurements
are consistent with the previous data from \citet{luh09fu}.
A comparison of the astrometry between 2008 and 2010 indicates that
FU~Tau~A and B share the same proper motion at a level of
$\sim5$~mas~yr$^{-1}$, which further supports the membership of these objects
in the same star-forming population, either as components of a binary system
or as unrelated Taurus members that are seen in projection near each other
\citep{luh09fu}.

\subsection{Reduction and PSF Analysis of Gemini Images}
\label{sec:gemini}

Our new Gemini AO images of 2M~J044144 and
2M~J044145 were reduced and analyzed in the same manner
as the first epoch from \citet{tod10} except that the data were
corrected for distortion with an IDL script provided by Chad Trujillo.
The first epoch data were reprocessed with this distortion correction as well.
The resulting measurements of separations and position angles between
the components of 2M~J044144~A/B and 2M~J044145~A/B
are presented in Table~\ref{tab:ao}. The relative positions of
2M~J044144~A and B from WFPC2 are also included \citep{tod10}.
Any errors that may be present in the distortion correction for the AO images
are not accounted for in the uncertainties in Table~\ref{tab:ao},
but they are only relevant to the measurement of relative astrometry across
large distances on the detector array, such as between
2M~J044144~A and 2M~J044145~A.
In addition, because each of the four objects was observed near the
same location on the array between the two epochs, a comparison of
the relative astrometry between epochs should not be affected
by errors in the distortion correction.

In \citet{tod10}, we demonstrated that 2M~J044144~A/B maintained
similar relative positions between the WFPC2 observation and the first
epoch of AO data, indicating that
the two objects share similar motions, and that the candidate secondary
is not a field star. Our second AO epoch
further supports this result, as shown in Table~\ref{tab:ao}. Meanwhile,
the relative positions of 2M~J044145~A/B and the relative
positions of 2M~J044144~A/2M~J044145~A
also were unchanged within the uncertainties between the two AO epochs.
For the first of these two pairs, if one component was motionless while
the other had the same proper motion as the nearest group of Taurus
members \citep[group V,][]{luh09tau}, then their separation and position angle
would change by $0\farcs006$ and $6\fdg4$, respectively, which is
inconsistent with our measurements. Thus, 2M~J044145~B shares
the same motion as the primary, and hence is a Taurus member.
Our astrometry is not sufficiently accurate to distinguish between
a binary system and a pair of unrelated Taurus members that have a small
projected separation, but the latter is very unlikely given the low 
stellar density in Taurus. For 2M~J044144~A and 2M~J044145~A, 
the separation and position angle would change by $0\farcs005$ and
$0\fdg12$, respectively, if one was motionless while the other exhibited
the proper motion of the nearest Taurus group. This relative motion is
not detectable in our data because the position angle error is dominated
by the uncertainty in the position angle of the camera.
However, both objects are already known to be members of
Taurus based on spectroscopy \citep{luh06tau,luh09tau,kra09}.

\section{Candidate Companions}
\label{sec:cand}

\subsection{Color-Magnitude Diagrams}

\begin{figure}
\epsscale{1}
\plotone{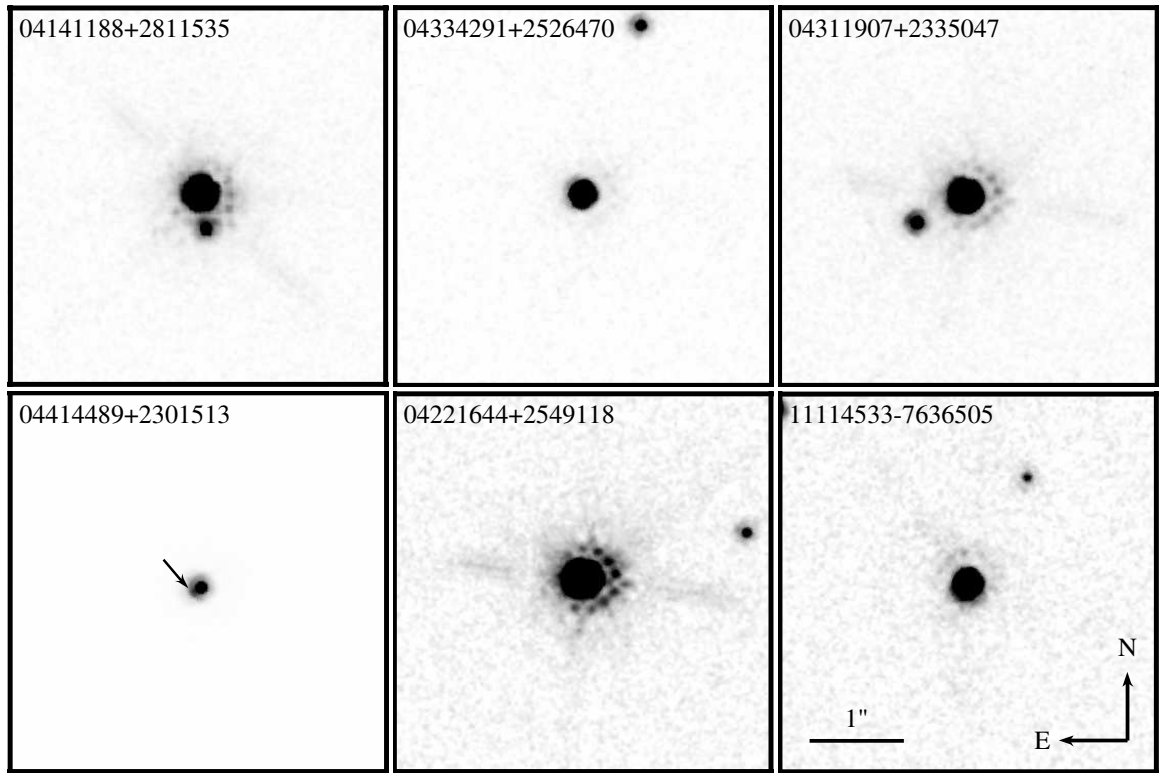}
\caption{
WFPC2 F791W images of young late-type members of Taurus and Chamaeleon~I
that have resolved candidate companions within $2\arcsec$. 
The photometry of the object near 2M~J044144 is consistent
with that expected for a member of Taurus while the remaining candidates
are probably field stars, as shown in Figure~\ref{fig:iz}.
PSF-subtracted images of the companion to 2M~J044144 were
presented by \citet{tod10}.
}
\label{fig:wide}
\end{figure}

\begin{figure}
\epsscale{1}
\plotone{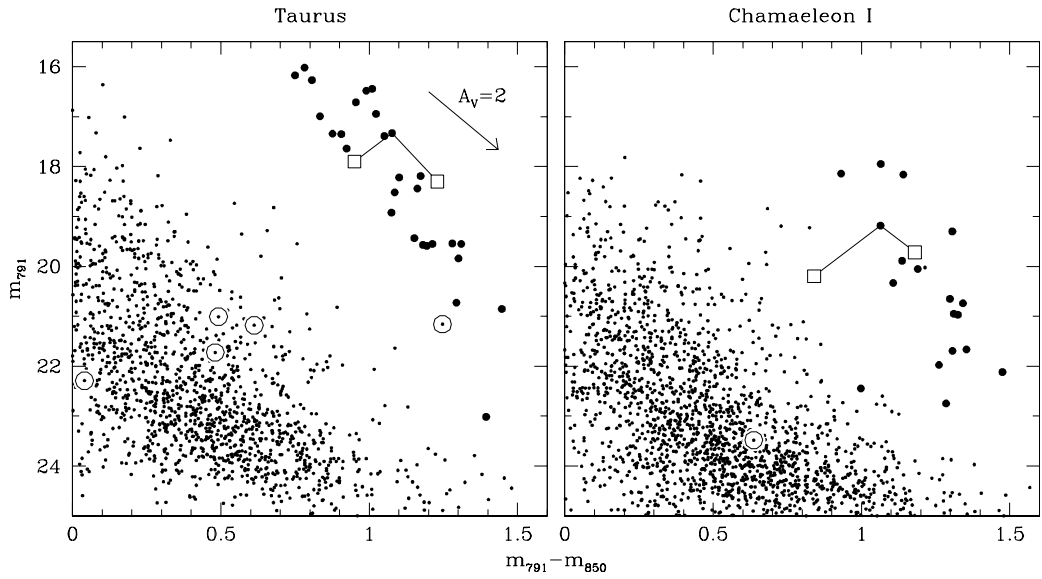}
\caption{
Color-magnitude diagrams constructed from WFPC2 images of late-type members
of the Taurus and Chamaeleon I star-forming regions.
We show the known members of these regions that are within the images
and are not saturated (large filled circles).
Among the remaining point sources (points), we indicate
the ones that are within $2\arcsec$ of a known member (circles).
One of these candidate companions appears within the Taurus sequence
and was confirmed as a member based on its proper motion
\citep[2M~J044144,][]{tod10}
while the other candidates are probably field stars based on their locations
below the sequences of known members.
We also plot the positions of the components of the partially resolved
binaries from Table~\ref{tab:bin} and Figure~\ref{fig:close} (squares).
The secondary for the pair in Chamaeleon~I is bluer than expected for
a cluster member, but its photometry is uncertain.
}
\label{fig:iz}
\end{figure}

Our WFPC2 images have detected point sources within a few arcseconds of
several of the primaries in Taurus and Chamaeleon~I. 
Images of the candidate companions with separations less than $2\arcsec$
are shown in Figure~\ref{fig:wide}. To assess the
companionship of these objects, we can check whether they have the colors
and magnitudes expected for members of these star-forming regions.
To do this, we have constructed in Figure~\ref{fig:iz} color-magnitude
diagrams for all unsaturated point sources in the WFPC2 images
of Taurus and Chamaeleon~I. Some of the Taurus primaries were saturated in
one of the bands and thus are absent from Figure~\ref{fig:iz}.
As noted in Section~\ref{sec:sample}, unsaturated photometry is available for
three additional members of Chamaeleon~I beyond the 15 low-mass primaries
that were targeted by WFPC2. The known members of these regions form
sequences that are well-separated from most field stars.
One exception is OTS~32, which appears below the sequence for
Chamaeleon~I. The subluminous appearance of this star has been observed
previously and probably indicates that it is seen in scattered light
\citep{luh08cha2}.

In Figure~\ref{fig:iz}, we have circled the candidate companions that are
within $2\arcsec$ of the primaries. One of these candidate companions,
2M~J044144~B, appears within the sequence of Taurus members
and is a likely companion \citep{tod10}. The remaining five candidates
fall below the cluster sequences and thus are probably field stars.
Among objects beyond $2\arcsec$ from the primaries (uncircled points), 
one appears within the Chamaeleon~I sequence, but its other optical and IR
colors are inconsistent with a cool object and suggest that it is a field star.
We have also marked in the color-magnitude diagrams the components 
of the two partially resolved binaries from Figure~\ref{fig:close}.
Both components of 2MASS J04221332+1934392 appear within the Taurus
sequence but the fainter component of ISO~217 is below the Chamaeleon~I
sequence. The anomalously blue color of ISO~217~B could indicate that it is
background field star or may result from the large uncertainties in its
photometry.
We note that all of the candidate companions from \citet{kra12} that
are in our WFPC2 images appear below the member sequence in
Figure~\ref{fig:iz}, and hence are probably field stars. 
The primaries for these candidates consist of 2MASS J04152409+2910434,
2MASS J04221644+2549118 (CFHT-14), 2MASS J04302365+2359129 (CFHT-16),
2MASS J04311907+2335047, and 2MASS J04334291+2526470.

\subsection{Probability of Companionship}

We now examine the probability of companionship for the three candidates
that have photometry consistent with membership in Taurus and Chamaeleon~I
(marginally consistent in the case of ISO~217~B). 
\citet{tod10} considered the companionship of 2M~J044144~B,
which has a separation of $0\farcs105$ from its primary.
As explained in that study, the probability that a field star with the
color and magnitude of a Taurus member would appear within $0\farcs1$
of any of the 32 Taurus primaries in our survey is $\sim10^{-5}$
(the astrometry from Section~\ref{sec:gemini} also indicates that 2M~J044144~B
is a Taurus member rather than a field star).
Given that 2MASS~J04221332+1934392~B has a separation of $0\farcs05$, the
probability that it is a field star is even lower.
Although ISO~217~B is below the cluster sequence for Chamaeleon~I, the
probability that it is a field star is not significantly higher than
these values in Taurus because the surface density of field stars is low
in its vicinity of the color-magnitude diagram in Figure~\ref{fig:iz}.
The unresolved spectroscopy of ISO~217~A+B from \citet{luh04cha} provides
additional constraints on the nature of the candidate secondary.
If ISO~217~B is a field star, it likely would be a reddened early-type
star or K giant. Since ISO~217~B contributes roughly 1/3 of the optical
flux of the pair, a field star of this kind probably would have been noticeable
in the unresolved spectrum. For instance, the molecular bands of the primary
would appear heavily diluted and veiled by the relatively featureless
continuum of a warmer field star. Finally, because of the low stellar densities
in Taurus and Chamaeleon~I, the components of these three pairs are
unlikely to be unrelated clusters members that have small projected
separations \citep{luh04bin,luh09fu}. Therefore, we conclude that
2M~J044144 and 2MASS~J04221332+1934392 likely comprise binary
systems. For ISO~217, we probably have detected either a binary companion or
a jet (see next section).

\subsection{Properties of Candidate Companions}

In \citet{tod10}, we estimated the physical properties for one of the three
companions that we have uncovered with WFPC2, 2M~J044144 B.
Its projected separation of $0\farcs105$ corresponds to 15~AU at the distance
of Taurus and its WFPC2 fluxes imply a mass of 5--10~$M_{\rm Jup}$ based on
theoretical evolutionary models. We now examine the properties of the
other two candidates, 2MASS~J04221332+1934392~B and ISO~217~B.
Their projected separations of $0\farcs051$ and $0\farcs031$ correspond to
7 and 5~AU, respectively, at the distances of Taurus and Chamaeleon~I.
As an unresolved pair, ISO~217~A+B exhibits strong H$\alpha$ emission
\citep{luh04cha,muz05}, forbidden emission lines \citep{sch06}, and
mid-IR excess emission \citep{apa05,luh05frac,luh08cha1}, indicating the
presence of active accretion and a circumstellar disk around at least one
component. In fact, a jet has been detected from ISO~217 \citep{whe09}, which
has a similar position angle as our candidate companion. Thus, it is possible
that we have detected emission from this jet rather than a companion,
although our filters should encompass little line emission from a jet, and
the brighter lobe of the jet is on the opposite side of the primary from
our candidate companion. We treat ISO~217 as a binary for the purposes
of this study.
Meanwhile, the components of 2MASS~J04221332+1934392 probably have roughly
similar masses given that the flux ratio at F850LP is near unity.
Neither component appears to have a close-in circumstellar disk 
based on the absence of mid-IR excess emission \citep{luh10tau}.

\subsection{Binary Statistics}
\label{sec:stat}

To characterize the multiplicity of low-mass stars and brown dwarfs
in Taurus and Chamaeleon~I, we combine the results from our survey
with those from previous high-resolution images in these regions.
The latter were collected with WFPC2 \citep{kra06}, Keck speckle
imaging \citep{kon07}, and Keck AO imaging \citep{kra12}
in Taurus and with WFPC2 \citep{neu02}, the Advanced Camera for Surveys
on {\it Hubble} \citep{luh07cha}, and AO at the Very Large
Telescope \citep{ahm07,laf08} in Chamaeleon~I.
For comparison to these two regions, we also have compiled binary
data measured for late-type members of the Upper Sco association
\citep[$\tau\sim11$~Myr,][]{pec12}
with WFPC2 and Keck AO \citep{kra05,bil11,kra12}\footnote{We have omitted 
USco~CTIO~132 and USco~CTIO~137  since they appear to be field dwarfs 
\citep[][K. Luhman, in prep]{muz03}.}.
We consider data for primaries with spectral types of $\geq$M4 
($\lesssim0.3$~$M_\odot$) using the classifications adopted by
\citet{luh08cha} and \citet{luh10tau,luh12u}.
For LH~0419+15, which was observed by \citet{kra06}, we adopt a type of M6
(K. Luhman, in preparation). The resulting samples contain 85, 66, and 50
primaries in Taurus, Chamaeleon~I, and Upper Sco, respectively, and are
compiled in Table~\ref{tab:comp}. 
These regions have similar distances ($d\sim150$~pc) and
have been observed with similar angular and mass detection limits,
which allows a direct comparison of their data.

\begin{figure}
\epsscale{1}
\plotone{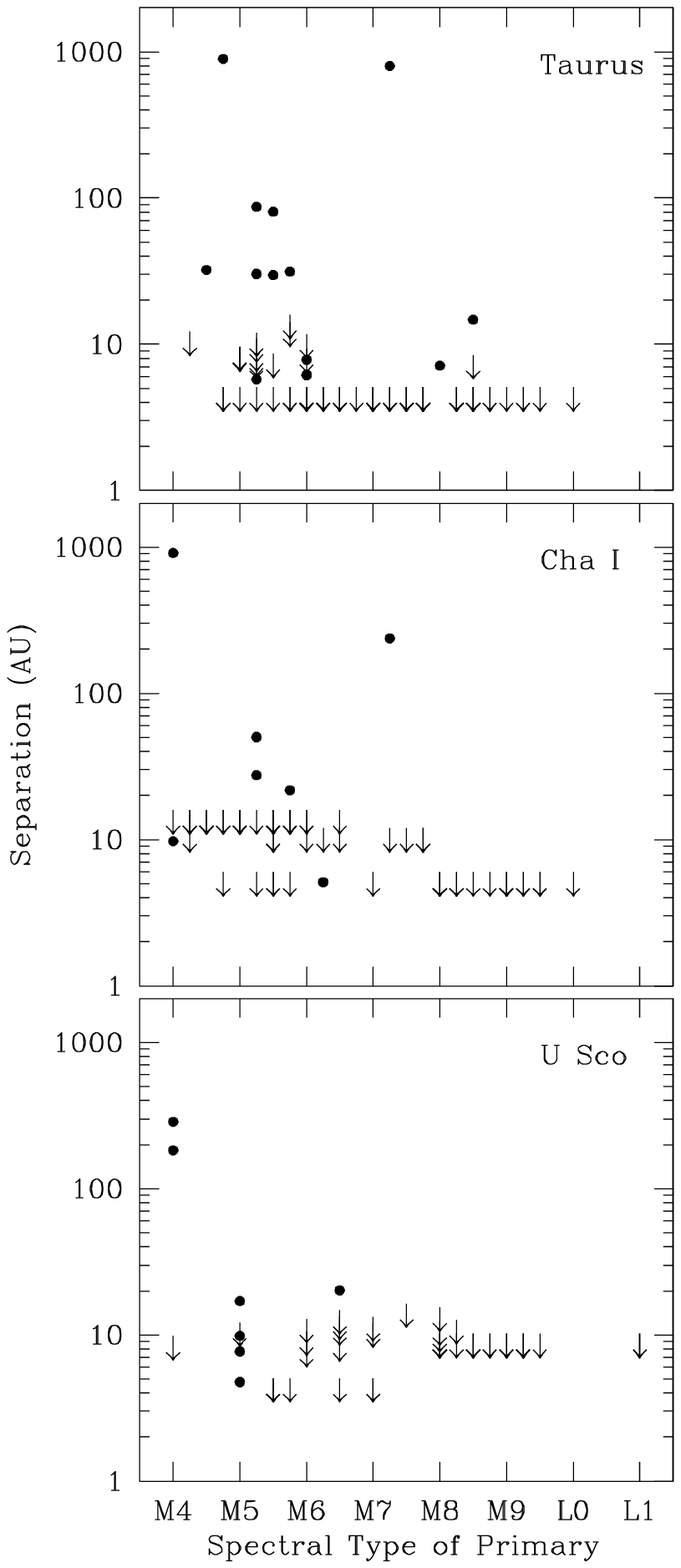}
\caption{
Binary measurements for late-type members of Taurus, Chamaeleon~I,
and Upper Sco \citep[][this work]{neu02,kra05,kra06,kon07,laf08,bil11,kra12}.
We show the projected separations of resolved binaries (points) and
the detection limits for unresolved sources (arrows) as a function
of the spectral type of the primary.
}
\label{fig:bin}
\end{figure}

In Table~\ref{tab:fraction}, we present the fractions of primaries in 
the high-resolution imaging surveys of Taurus, Chamaeleon~I, and Upper Sco
for which probable companions have been detected at $>$10~AU, which is 
the smallest separation that all of these surveys reached.
Separate fractions are shown for M4--M6 (0.1--0.3~$M_\odot$) and $>$M6
($\lesssim0.1$~$M_\odot$) so that we can examine the dependence of the binary
fractions on spectral type, and hence stellar mass.
To illustrate the distribution of separations and how it varies with
spectral type of the primary, we plot the separations of resolved binaries
and the separation limits for equal-magnitude pairs among the unresolved
primaries versus spectral type in Figure~\ref{fig:bin}.
For this diagram, we have adopted separation limits
of $0\farcs03$ for the PC, Advanced Camera, and speckle data,
$0\farcs06$ for the WFC and NICMOS data,
$0\farcs08$ for the AO data from the Very Large Telescope,
and the FWHM of the Keck AO data. We note that some AO data are capable
of detecting binaries with separations that are smaller than the FWHM
\citep{kra12}.

\section{Discussion}
\label{sec:disc}

Taurus, Chamaeleon~I, and Upper Sco contain the largest samples of young
low-mass stars and brown dwarfs that have been imaged
at high resolution, providing the best available statistical
constraints on low-mass multiplicity at ages of $\lesssim10$~Myr. 
We first examine the dependence of the binary fractions in
Table~\ref{tab:fraction} on spectral type of the primary.
For each of the three regions, the wide binary fraction ($>$10~AU) is
significantly lower at $>$M6 than at M4--M6.
Trends of this kind have been detected previously in
subsets of the data we have compiled \citep{kra05,kra06,bil11,kra12}, in
samples of members of Taurus, Chamaeleon~I, and Upper Sco at
higher masses \citep{laf08,kra09}, and in the solar neighborhood
\citep[][references therein]{bur07ppv}.
Given the youth and low density of the regions in question, particularly
Taurus, this dependence on spectral type is very likely primordial rather
than due to dynamical interactions among members of each region.

We also can examine the data in Table~\ref{tab:fraction} for differences
among Taurus, Chamaeleon~I, and Upper Sco.
For each range of spectral types, the binary fractions do not show any
statistically significant differences between the regions.
Comparing these binary fractions to data for field stars
and brown dwarfs is more problematic because it is difficult to ensure
that young and old samples encompass the same ranges of primary masses,
and because a given young cluster may not represent the predominant
star-forming environment for the field.
Nevertheless, it is useful to compare the frequency of the widest brown
dwarf binaries. Taurus and Chamaeleon~I each contain one known binary
brown dwarf with a separation greater than 100~AU \citep{luh04bin,luh09fu}.
Although the Upper Sco sample that we have defined for
Table~\ref{tab:fraction} does not have any pairs wider than 100~AU,
a few examples have been found among other brown dwarfs in the association
\citep{all06,jay06,clo07,luh07oph,bej08}. Thus, the data for
Taurus, Chamaeleon, and Upper Sco indicate a binary fraction of a few percent
for these wide binary brown dwarfs. In comparison, only one pair of this
kind has been found among the several hundred known late-L and T dwarfs
in the field \citep{burn10,sch10}.
This implies that dynamical interactions with cluster members or field
stars at $>$10~Myr disrupt the widest binary brown dwarfs, or that
most field brown dwarfs are born under conditions different from those in
Taurus, Chamaeleon~I, and Upper Sco, perhaps in denser clusters where
very wide binaries are disrupted or prevented from forming.

\acknowledgements

We thank Allison Youngblood, Steven Mohammed, Ijeoma Ekeh, and Jaclyn Payne
for assistance with the data analysis.
We acknowledge support from grant AST-0544588 from the National Science
Foundation and grants GO-11203, GO-11204, and GO-11983 from the
Space Telescope Science Institute. D.A. also acknowledges support from grant
NNX11AG57G from the NASA Origins of Solar Systems program.
The Gemini data were obtained
through program GN-2011A-Q-10. Gemini Observatory is operated
by AURA under a cooperative agreement with the NSF on behalf of the
Gemini partnership: the NSF (United States), the Particle Physics and
Astronomy Research Council (United Kingdom), the National Research
Council (Canada), CONICYT (Chile), the Australian Research Council
(Australia), CNPq (Brazil) and CONICET (Argentina).
The Center for Exoplanets and Habitable Worlds is supported by the
Pennsylvania State University, the Eberly College of Science, and the
Pennsylvania Space Grant Consortium. MultiDrizzle is a product of the
Space Telescope Science Institute, which is operated by AURA for NASA.

\clearpage

\begin{deluxetable}{llllll}
\tabletypesize{\scriptsize}
\tablewidth{0pt}
\tablecaption{Members of Taurus and Chamaeleon I in the WFPC2 images\label{tab:st}}
\tablehead{
\colhead{2MASS\tablenotemark{a}} &
\colhead{Other Name} &
\colhead{Spectral Type} &
\colhead{m$_{791}$\tablenotemark{b}} &
\colhead{m$_{850}$\tablenotemark{b}} &
\colhead{Date}}
\startdata
J04141188+2811535 &    \nodata &  M6.25 & 17.35 & 16.45 & 2008 Nov 8 \\
J04152409+2910434 &    \nodata &     M7 & 18.22 & 17.12 & 2008 Aug 1 \\
J04161885+2752155 &    \nodata &  M6.25 & 16.72 & 15.76 & 2007 Sep 15 \\
J04201611+2821325 &    \nodata &   M6.5 & 17.64 & 16.72 & 2007 Oct 24 \\
J04215450+2652315 &    \nodata &   M8.5 & 20.85 & 19.40 & 2008 Jul 31 \\
J04221332+1934392A+B &    \nodata &     M8 & 17.33\tablenotemark{c} & 16.25\tablenotemark{c} & 2007 Aug 6 \\
J04221644+2549118 &    \nodata &  M7.75 & 17.39 & 16.34 & 2008 Aug 1 \\
J04242090+2630511 &    \nodata &   M6.5 & 17.35 & 16.47 & 2008 Aug 1 \\
J04263055+2443558 &    \nodata &  M8.75 & 19.54 & 18.26 & 2008 Aug 9 \\
J04270739+2215037 &    \nodata &  M6.75 & 16.02 & 15.23 & 2008 Aug 2 \\
J04274538+2357243 &    \nodata &  M8.25 & 19.59 & 18.39 & 2008 Aug 1 \\
J04290068+2755033 &    \nodata &  M8.25 & 18.45 & 17.28 & 2008 Aug 15 \\
J04302365+2359129 &    \nodata &  M8.25 & 19.57 & 18.39 & 2008 Aug 6 \\
J04311907+2335047 &    \nodata &  M7.75 & 18.19 & 17.02 & 2008 Aug 6 \\
J04312669+2703188 &    \nodata &   M7.5 & 19.55 & 18.34 & 2008 Aug 10 \\
J04320329+2528078 &    \nodata &  M6.25 & 15.51 &   sat & 2008 Aug 15 \\
J04322329+2403013 &    \nodata &  M7.75 & 16.48 & 15.49 & 2008 Sep 7 \\
J04334291+2526470 &    \nodata &  M8.75 & 19.55 & 18.24 & 2008 Aug 12 \\
J04350850+2311398 &    \nodata &     M6 & 16.27 & 15.46 & 2007 Oct 6 \\
J04354526+2737130 &    \nodata &  M9.25 & 19.84 & 18.54 & 2008 Aug 4 \\
J04361030+2159364 &    \nodata &   M8.5 & 19.43 & 18.28 & 2008 Aug 15 \\
J04373705+2331080 &    \nodata &     L0 & 23.01 & 21.62 & 2007 Aug 25 \\
J04385871+2323595 &    \nodata &   M6.5 & 16.07 &   sat & 2008 Aug 4 \\
J04390396+2544264 &    \nodata &  M7.25 & 16.95 & 15.93 & 2008 Aug 6 \\
J04390637+2334179 &    \nodata &   M7.5 & 15.71 &   sat & 2008 Aug 18 \\
J04400067+2358211 &    \nodata &     M6 & 16.17 & 15.42 & 2007 Aug 24 \\
J04414489+2301513A &    \nodata &   M8.5 & 18.93 & 17.85 & 2008 Aug 20 \\
J04414489+2301513B &    \nodata &   \nodata & 21.16 & 19.91 & 2008 Aug 20 \\
J04414825+2534304 &    \nodata &  M7.75 & 18.52 & 17.44 & 2007 Oct 27 \\
J04442713+2512164 &     IRAS 04414+2506   &  M7.25 & 16.45 & 15.44 & 2007 Aug 19 \\
J04484189+1703374 &    \nodata &     M7 &   sat & 16.59 & 2007 Aug 8 \\
J04552333+3027366 &    \nodata &  M6.25 & 16.99 & 16.16 & 2007 Aug 26 \\
J04574903+3015195 &    \nodata &  M9.25 & 20.73 & 19.43 & 2007 Aug 26 \\
J11011926-7732383A &    \nodata &  M7.25 & 18.16 & 17.02 & 2009 Apr 30 \\
J11011926-7732383B &    \nodata &  M8.25 & 19.30 & 18.00 & 2009 Apr 30 \\
J11020610-7718079 &    \nodata &     M8 & 20.65 & 19.35 & 2009 Mar 16 \\
J11025374-7722561 &    \nodata &   M8.5 & 20.96 & 19.64 & 2009 Mar 4 \\
          \nodata & Cha J11062854-7618039 &     M9 & 21.98 & 20.72 & 2009 Apr 24 \\
          \nodata & Cha J11070768-7626326 &     L0 & 22.75 & 21.46 & 2009 Apr 24 \\
J11082570-7716396 &    \nodata &     M8 & 21.69 & 20.38 & 2009 Apr 23 \\
J11084952-7638443 &    \nodata &  M8.75 & 20.95 & 19.64 & 2009 Apr 21 \\
J11095215-7639128A+B &     ISO 217 A+B &  M6.25 & 19.18\tablenotemark{c} & 18.12\tablenotemark{c} & 2009 Apr 26 \\
J11095505-7632409 &       C1-2 & \nodata & 20.33 & 19.23 & 2009 Apr 23 \\
J11100336-7633111 &      OTS 32 &     M4 & 22.44 & 21.45 & 2009 Apr 23 \\
J11100658-7642486 &    \nodata &  M9.25 & 21.67 & 20.31 & 2009 Apr 24 \\
J11100934-7632178 &      OTS 44 &   M9.5 & 22.11 & 20.64 & 2009 Apr 23 \\
J11102226-7625138 &  CHSM 17173 &     M8 & 17.95 & 16.88 & 2009 Apr 27 \\
J11112249-7745427 &    \nodata &  M8.25 & 20.05 & 18.86 & 2009 Apr 24 \\
J11114533-7636505 &    \nodata &     M8 & 19.89 & 18.75 & 2009 Feb 27 \\
J11122250-7714512 &    \nodata &  M9.25 & 20.74 & 19.40 & 2009 Apr 24 \\
J11123099-7653342 &    \nodata &     M7 & 18.14 & 17.21 & 2009 Mar 3 \\
\enddata
\tablenotetext{a}{2MASS Point Source Catalog \citep{skr06}.}
\tablenotetext{b}{An entry of ``sat" indicates that the object was saturated
in this band.}
\tablenotetext{c}{This photometry applies to the combined flux from this
partially resolved binary (Figure~\ref{fig:close}).
The flux ratio for the components of the binary is in Table~\ref{tab:bin}.}
\end{deluxetable}

\begin{deluxetable}{lllll}
\tabletypesize{\scriptsize}
\tablewidth{0pt}
\tablecaption{Astrometry and Photometry for Partially Resolved
Binaries from WFPC2\label{tab:bin}}
\tablehead{
\colhead{Name} &
\colhead{$\rho$} &
\colhead{P.A.\tablenotemark{a}} &
\multicolumn{2}{c}{Flux Ratio\tablenotemark{b}} \\
\cline{4-5}
\colhead{} &
\colhead{(arcsec)} &
\colhead{(deg)}&
\colhead{F791W} &
\colhead{F850LP}}
\startdata
2MASS J04221332+1934392 & 0.051$\pm$0.003 & 316$\pm$4  & 0.69$\pm$0.03 & 0.89$\pm$0.03\\
ISO 217                 & 0.031$\pm$0.004 & 238$\pm$8  & 0.64$\pm$0.05 & 0.47$\pm$0.08\\
\enddata
\tablenotetext{a}{Position angle of the secondary relative to the primary.}
\tablenotetext{b}{The combined photometry of each binary is in
Table~\ref{tab:st}.}
\end{deluxetable}

\begin{deluxetable}{llc}
\tablecolumns{3}
\tabletypesize{\scriptsize}
\tablewidth{0pt}
\tablecaption{Astrometry for Components of 2MASS J04414489+2301513
and 2MASS J04414565+2301580\label{tab:ao}}
\tablehead{
\colhead{$\rho$} &
\colhead{P.A.\tablenotemark{a}} &
\colhead{Date} \\
\colhead{(arcsec)} &
\colhead{(deg)} &
\colhead{}}
\startdata
\cutinhead{2MASS J04414489+2301513~A and B}
0.105$\pm$0.004 & 120.4$\pm$2.2 & 2008 Aug 20 \\
0.105$\pm$0.005 & 120.7$\pm$2.6 & 2009 Oct 13 \\
0.104$\pm$0.006 & 122.8$\pm$3.0 & 2011 Feb 13 \\
\cutinhead{2MASS J04414565+2301580~A and B}
0.226$\pm$0.004 & 84.8$\pm$1.0 & 2009 Oct 13 \\
0.230$\pm$0.006 & 84.6$\pm$1.4 & 2011 Feb 13 \\
\cutinhead{2MASS J04414565+2301580~A and 2MASS J04414489+2301513~A}
12.325$\pm$0.004 & 238.0$\pm$0.1 & 2009 Oct 13 \\
12.325$\pm$0.006 & 237.9$\pm$0.1 & 2011 Feb 13 \\
\enddata
\tablenotetext{a}{Position angle of the secondary relative to the primary.}
\end{deluxetable}

\begin{deluxetable}{lllll}
\tabletypesize{\scriptsize}
\tablewidth{0pt}
\tablecaption{Late-type Targets of Multiplicity Surveys in Taurus, Cha~I, and Upper Sco\label{tab:comp}}
\tablehead{
\colhead{2MASS\tablenotemark{a}} &
\colhead{Other Name} &
\colhead{Spectral Type} &
\colhead{$\rho$\tablenotemark{b}} &
\colhead{Reference}\\
\colhead{} &
\colhead{} &
\colhead{} &
\colhead{(arcsec)} &
\colhead{}}
\startdata
J04141188+2811535 &    \nodata &  M6.25 & $<$0.03,$<$0.053 & 1,2 \\
J04151471+2800096 &          KPNO 1   &   M8.5 & $<$0.03 & 3 \\
J04152409+2910434 &    \nodata &     M7 & $<$0.03,$<$0.051 & 1,2 \\
J04161210+2756385 &    \nodata &  M4.75 & $<$0.03 & 4 \\
J04161885+2752155 &    \nodata &  M6.25 & $<$0.03,$<$0.078 & 1,2 \\
\enddata
\tablecomments{
This table is available in its entirety in a machine-readable form in the
online journal. A portion is shown here for guidance regarding its form and
content.}
\tablenotetext{a}{2MASS Point Source Catalog \citep{skr06}.}
\tablenotetext{b}{This column contains the projected separations of
resolved binaries and the detection limits for unresolved sources.
The adopted limits are described in Section~\ref{sec:stat}.}
\tablerefs{
(1) this work;
(2) \citet{kra12};
(3) \citet{kra06};
(4) \citet{kon07};
(5) \citet{luh09fu};
(6) \citet{tod10};
(7) \citet{luh04bin};
(8) \citet{laf08};
(9) \citet{ahm07};
(10) \citet{luh07cha};
(11) \citet{neu02};
(12) \citet{bil11};
(13) \citet{kra05}.
}
\end{deluxetable}

\begin{deluxetable}{lllll}
\tabletypesize{\scriptsize}
\tablewidth{0pt}
\tablecaption{Low-mass Binary Fractions ($>$10~AU) in Young Regions\label{tab:fraction}}
\tablehead{
\colhead{Spectral Type} &
\colhead{Taurus\tablenotemark{a}} &
\colhead{Cha I\tablenotemark{b}} &
\colhead{U Sco\tablenotemark{c}} &
\colhead{Total}}
\startdata
M4--M6 & 7/39 = $0.18^{+0.08}_{-0.04}$ &  4/39 = $0.1^{+0.07}_{-0.03}$ &   3/15 = $0.2^{+0.14}_{-0.06}$ & 14/93 = $0.15^{+0.05}_{-0.03}$ \\
$>$M6  &  2/46 = $0.04^{+0.05}_{-0.01}$ &  1/27 = $0.04^{+0.07}_{-0.01}$ &   1/35 = $0.03^{+0.06}_{-0.01}$ & 4/108 = $0.04^{+0.03}_{-0.01}$ \\
\enddata
\tablenotetext{a}{\citet{kra06}, \citet{kon07}, \citet{kra12}, and this work.}
\tablenotetext{b}{\citet{neu02}, \citet{ahm07}, \citet{luh07cha},
\citet{laf08}, and this work.}
\tablenotetext{c}{\citet{kra05}, \citet{bil11}, and \citet{kra12}.}
\end{deluxetable}

\end{document}